\documentclass{article} 
\usepackage{nips15submit_e,times}
\usepackage{url}
\usepackage{graphicx,float}
\usepackage{amssymb,amsmath,amsthm}
\usepackage{rotating}
\newcommand{\ignore}[1]{}

\title{Community dynamics in connected \\ time-dependent multilayer networks}

\ignore{
\author{
Marco Cristoforetti\\
Fondazione Bruno Kessler\\ 
Trento (Italy)\\
\texttt{mcristofo@fbk.eu} \\
\And
Marco Guerini\\
Fondazione Bruno Kessler\\ 
Trento (Italy)\\
\texttt{guerini@fbk.eu} \\
\And
Giuseppe Jurman\\
Fondazione Bruno Kessler\\ 
Trento (Italy)\\
\texttt{jurman@fbk.eu} \\
\And
Cesare Furlanello\\
Fondazione Bruno Kessler\\ 
Trento (Italy)\\
\texttt{furlan@fbk.eu} 
}
}

\author{
Marco Cristoforetti$\qquad$Marco Guerini$\qquad$Giuseppe Jurman\thanks{Corresponding author}$\qquad$Cesare Furlanello\\
Fondazione Bruno Kessler\\
Trento (Italy)\\
\texttt{\{mcristofo,guerini,jurman,furlan\}@fbk.eu}
}

\nipsfinalcopy 

\begin{document}
\maketitle

\begin{abstract}
Different strategies have been considered to extract information from social media about how similarly people react to the same news or event. In this context, a powerful method is offered by the application of graph techniques to the contents produced by social network users. In particular, large events typically attract enough content traffic along time to enable an analysis that explicitly models a dependence from the time dimension. Here we demonstrate how it is possible to extend the application of community detection strategies in complex networks to the case of time-dependent multilayer networks, whenever the connection between consecutive time layers is non-trivial. We apply the method to 400K Twitter post related to the Expo event held in Milan (Italy) between May and October 2015.
\end{abstract}

\section{Introduction}
\label{sec:intro}
Understanding human social interaction and transmission of information requires the application of interdisciplinary strategies mixing techniques from psychology to statistics and algorithmic approaches such as those developed in graph theory. Since the social media explosion, a lot of effort has been put in trying to understand how information is mediated on these channels and how these data can improve the general understanding of social behaviour~\cite{centola2010spread}.

Complex network theory represents the elective framework for application to data flows from social media and in particular for the Twitter microblogging platform~\cite{java2007we,wu2011says,myers2012information,bakshy2011everyone,omodei2015characterizing,beguerisse2014interest,gargiulo2014topology}.
A major issue of interest is identifying birth, growth and decline of communities between members of a social network. Detection of communities is not trivial, and it is even harder to track the time dynamics of the community throughout its whole lifespan. This problem is not specific of social media related data, but it is indeed common to every situation where the behavior of a large group of people is observed and tentatively modelled. A classic example is community detection in parliamentary activity, where voting behavior can be used to group politicians in unofficial groups that fall outside the party ownership~\cite{porter2005network,moody2015portrait}.

In this paper we focus on a variation of the usual modularity approach in time-dependent networks~\cite{mucha2010community,kawadia2012sequential,darmon2013detecting}, trying to follow closely the evolution in time of the communities by an explicit modeling of the interlayer connectivity between consecutive time-layers with a non-trivial adjacency block-matrix.
We introduce the method via its first application to Twitter data about the Universal Exhibition Expo 2015, hosted in Milan (Italy) from May 1 to October 31.  We focus on data coming from the Twitter platform, connected to this main event, and generated between May 11 and October 18. Beyond being of the largest world events in 2016, Expo 2015 on Twitter is an interesting case study for our method because aside the core hashtag \#expo2015, different topics attracted the attention of the users and led to the formation of discussion communities.

\section{Expo2015 Twitter data}
\label{sec:data}
Twitter is particularly suitable to follow activity connected with a specific event. Most of the tweets are public and the official API released from the platform permits to collect the messages containing a given selection of words. In the case of Expo2015, we retrieved all the tweets containing ``\textit{expo2015}'' and ``\textit{noexpo}'' in the text.
From the overall collection of 1.4 million tweets, we present here the analysis on all messages in Italian from May 11 until October 18, covering five of the opening six months of the exhibition. This data subset counts more than 400K tweets from around 60K different users and 40K hashtags. Different networks can be defined from the dataset;  to start with, either hashtags or users can be used as vertices. As we study communities, in this study nodes are defined by the users, whose relationships are basically measured by analyzing retweet, reply or mention activities. We intend to follow topics and how tweeters are involved in discussing topics and we start by drawing an edge connecting two users if they share one or more hashtags. In order to have meaningful networks we remove ``\textit{expo2015}'' from the user hashtag list, which is the query keyword in the Twitter API and therefore it connects almost all tweets. Moreover, we filter out 11 generic hashtags that do not convey a specific message: ``\textit{expomilano}'', ``\textit{expo}'', ``\textit{milano}'', ``\textit{milan}'', ``\textit{expomilano2015}'', ``\textit{milanoexpo2015}'', ``\textit{expo2015milano}'', ``\textit{euexpo2015}'', ``\textit{e015}'', ``\textit{tim2go}'', ``\textit{news}''.

\section{Methods}
\label{sec:methods}

The aim of our method is modeling the evolution of communities in time-dependent multilayer networks. For the Twitter data related to Expo2015 described in Section~\ref{sec:data}, we are considering a time interval of slightly more than five months (21 weeks). In this section we outline the preprocessing of the dataset, covering the encoding of time related features and the definition an appropriate community detection algorithm. In particular, we introduce the notion of connected time-dependent network ($\mathcal{N}$) as underlying graph structure; among the available community detection algorithms, we chose the two-level InfoMap (\url{http://www.mapequation.org}) based on the optimization of the map equation~\cite{rosvall2008maps,rosvall2010map}.

\subsection{Connected time-dependent network - $\mathcal{N}$}
\label{ssec:time_nets}

Daily data were considered, for a total of 161 days. Smaller time-scales, as hours, are not optimal. Smaller time-scales, as hours, are not optimal because the activity during the night is negligible with respect to that on the day ($6.5\%$ from $23$pm to $5$am). Larger scales tends to flatten too much the dynamics and possibly hide interesting patterns. Networks are not multiplex, because users posted sparsely ($1063\pm 392$ users per day) and with different number of tweets during the observation period (an average of $4.3$ tweets per user, excluding professionals and bots, but with $59\%$ of the users that posted only once); therefore different nodes are found between networks at different timestamps. The connected time–dependent network $\mathcal{N}$  is defined by inter-layer edges in the same way as the intra-layer links: different nodes in consecutive layers are linked whenever they share one or more hashtags. In terms of adjacency matrix $\mathbf{A}$, the resulting $\mathcal{N}$ can be represented as a block diagonal matrix, where the blocks corresponds to the days, with first off-diagonal blocks representing the interactions between consecutive days.

\begin{figure}[!ht]
\begin{tabular}{cc}
\qquad$
\mathbf{A} = \left(
\begin{array}{cccc}
N_{t_1} & N_{t_1t_2} & 0 &0 \\
N_{t_1t_2} & N_{t_2} & 0 & 0 \\
0 & 0 & \ddots & N_{t_{n-1}t_n} \\
0 & 0 & N_{t_{n-1}t_n} & N_n
\end{array}
\right)
$
&
\qquad\raisebox{-11mm}{\includegraphics[width=.25\textwidth]{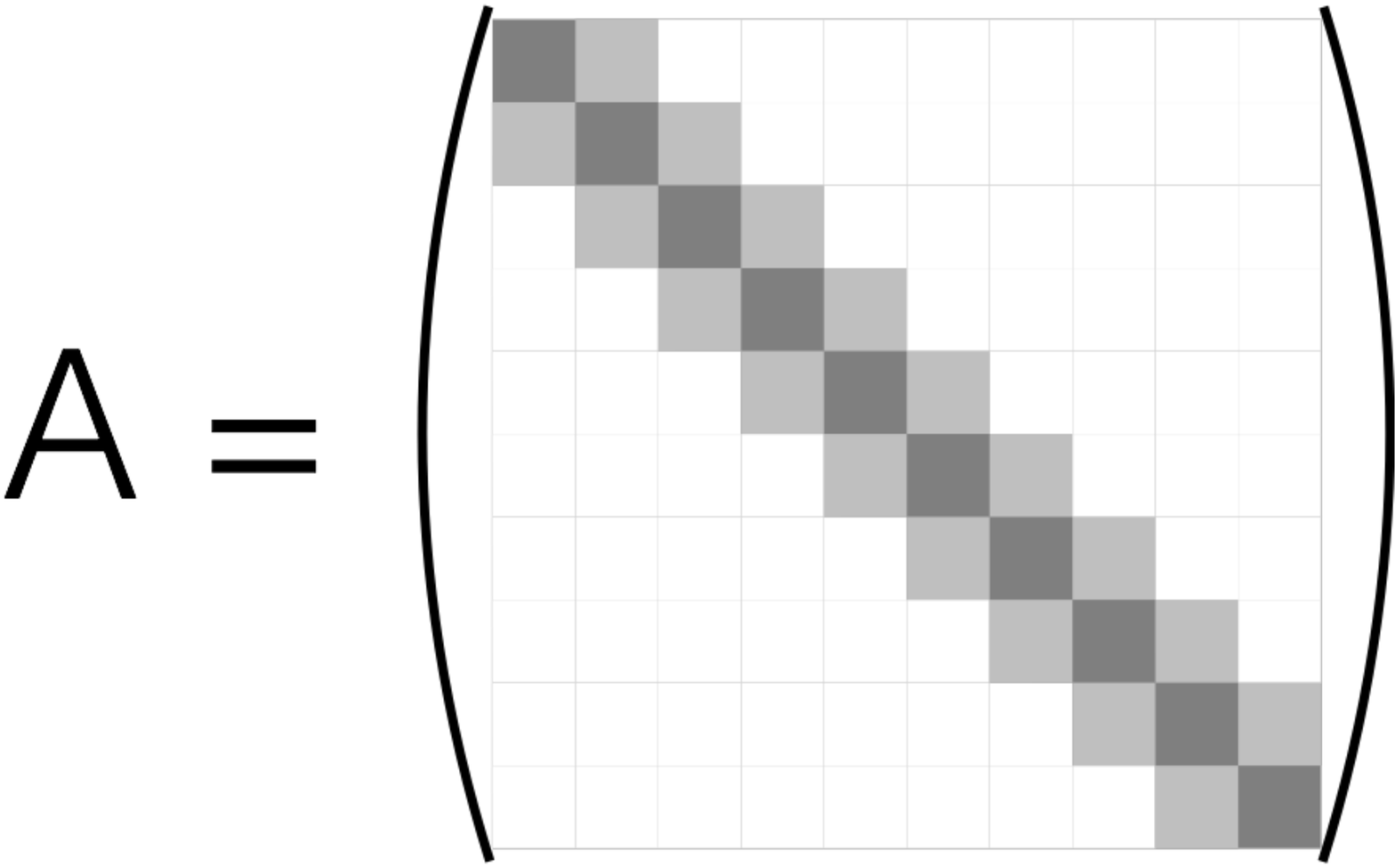}}
\end{tabular}
\label{fig:matrix}
\caption{Mathematical definition and graphical representation of the block adjacency matrix for $\mathcal{N}$}
\end{figure}

Throughout the entire time span of 161 days, about one thousand users tweeted at least one. 
About $60\%$ of these people posted only one tweet, yielding that only a few users are present on several network layers. In practice, bots and professional users posting daily several tweets would be the only nodes appearing in a multiplex networks built with the aforementioned rules.

\subsection{Community detection}
\label{sec:infomap}

In our methods, community detection is performed by using the optimization of the map equation (two level InfoMap algorithm)~\cite{rosvall2010map,rosvall2008maps}, based on information theory to look for the best partition of the network. The community structure is represented through a two-level nomenclature based on Huffman coding: the first level spots the communities in the network and the second picks up nodes in each community. According to the criterion, finding the best partition reads as minimizing the quantity of information needed to represent a random walk in the network. For a partition containing few intercommunity links, the walker will stay longer inside communities, therefore only the second level will be needed to describe its path, leading to a compact representation. As we aim to track the evolution of topics, this approach looks particularly appealing as it considers the flow of information in the network. Alternative community algorithms~\cite{fortunato2010community} were also considered, but they were deemed unfeasible either for theoretical (\textit{e.g.}, Louvain methods tends to join nodes in too large communities~\cite{fortunato2007resolution}) or computational (for the stochastic block model~\cite{yang2015scalable}) reasons.

\section{Results}
\label{sec:results}
\begin{figure}[!ht]
\begin{center}
\begin{tabular}{cc}
\includegraphics[width=0.5\textwidth]{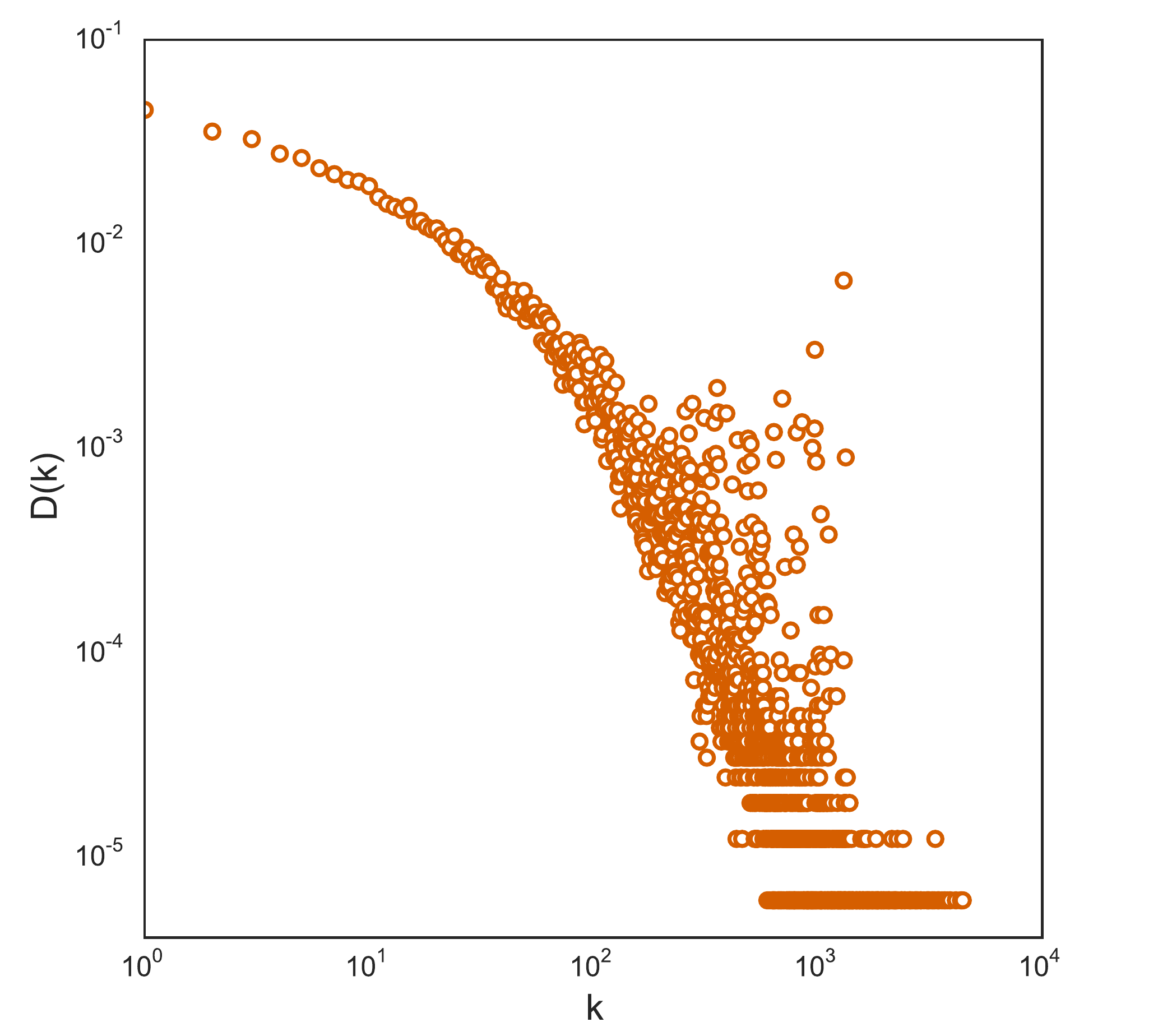} &
\includegraphics[width=0.5\textwidth]{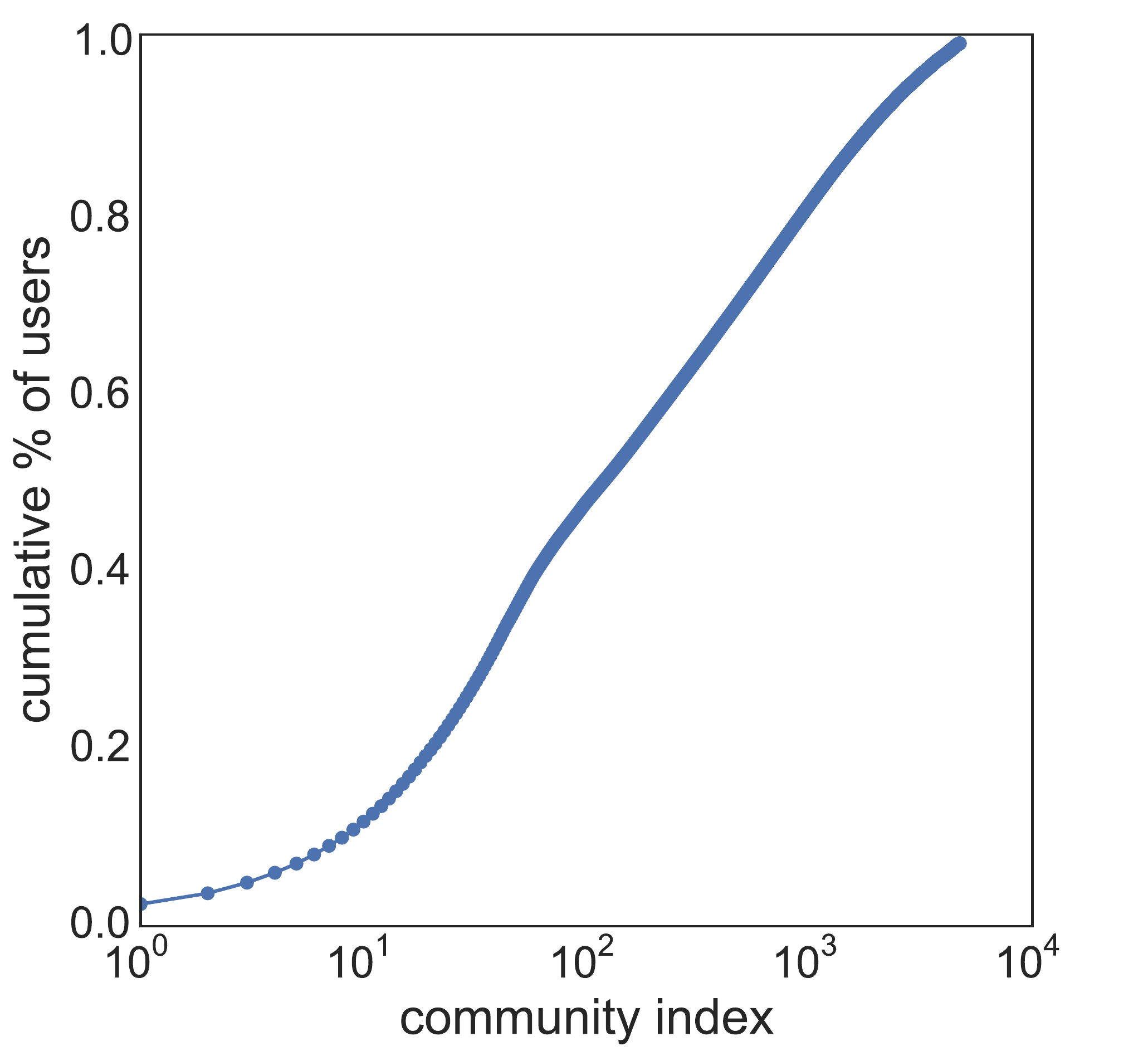} \\
(a) & (b)
\end{tabular}
\end{center}
\caption{(a) The degree distribution of $\mathcal{N}$. (b) Relation between the ranked communities and the corresponding normalized cumulative number of involved users.}
\label{fig:fig1}
\end{figure}

The connected time-dependent network $\mathcal{N}$ for the Expo2015 dataset is formed by $168.041$ nodes and $8.033.016$ edges.
Its degree distribution, shown in Fig.~\ref{fig:fig1}(a), does not follow the power-law trend, since a significant number of nodes have a large degree. 
This pattern is in part explained by the contribute of bots and professional users, as well as of highly common hashtags generating large cliques, e.g.  ``\textit{expottimisti}'' or ``\textit{food}''.

As a direct consequence, $\mathcal{N}$ has a quite large average local clustering coefficient:
\begin{displaymath}
\bar{C}=\frac{1}{N}\sum_{i=1}^N C_i=\frac{2|{e_{jk}:v_j,v_k\in N_i, e_{jk}\in E}|}{k_i(k_i-1)} = 0.76,
\end{displaymath}
where $e_{jk}$ indicates the edge between users $j$ and $k$.

As explained in Sec.~\ref{sec:infomap}, we apply the two-level map equation optimization for community detection over $\mathcal{N}$.
Operatively we consider $\mathcal{N}$ as a unique network over the full 161 days period, but in principle we could have considered shorter periods of time and we need to be sure that our results are not affected from the length of the frame selection. 
Moreover the independence of the results from the time interval also helps parallelizing the computation: the full network $\mathcal{N}$ is split in blocks which can be processed in parallel without affecting the results.
In order to test the independence hypothesis, a community detection analysis was run covering time frames of $2,\ 4,\ 8,\ 16,\ 32,\ 64,\ 128$ days and on the whole $\mathcal{N}$. 

\begin{figure}[!ht]
\begin{center}
\includegraphics[width=0.5\textwidth]{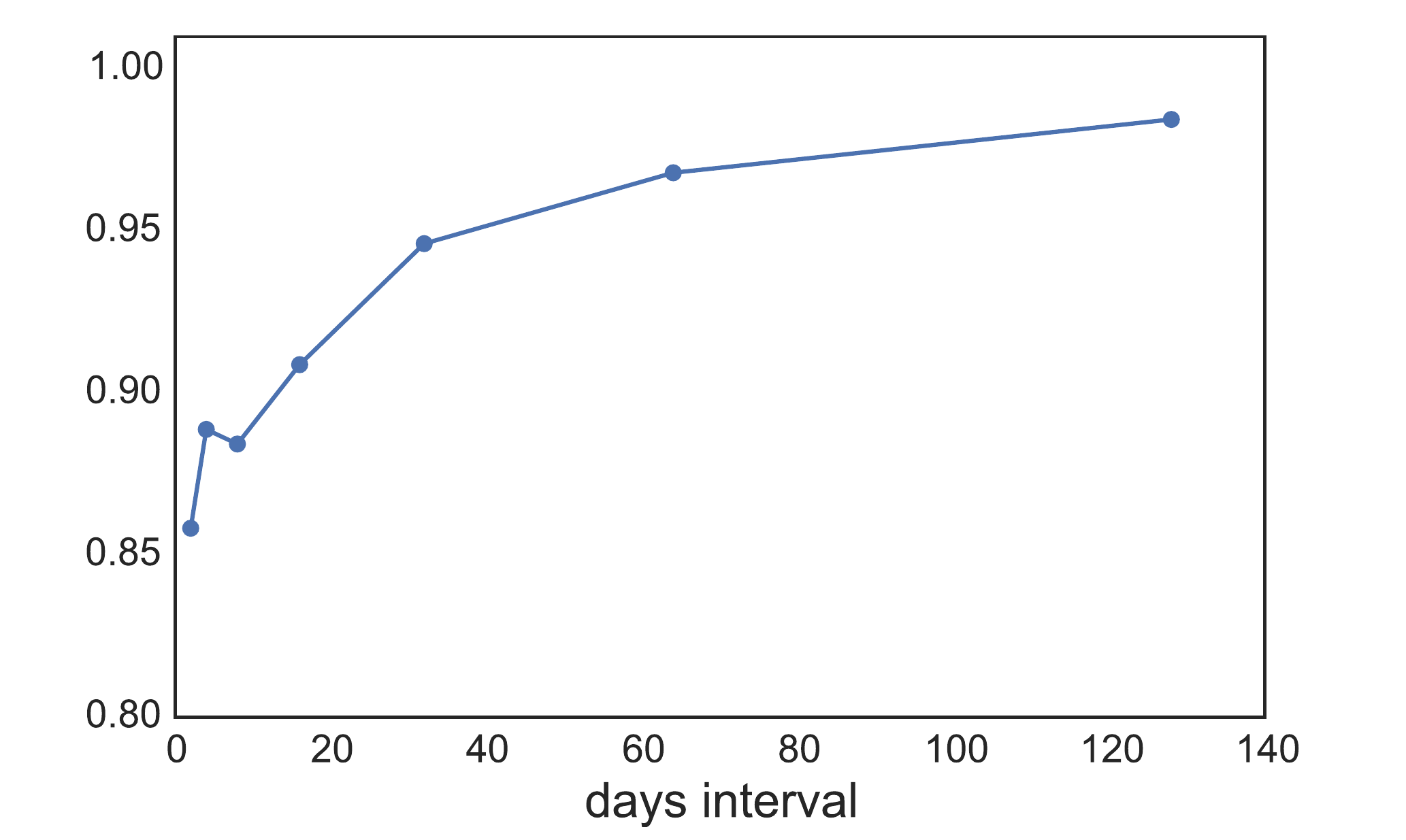}
\end{center}
\caption{Ratio between the number of communities identified in the first $n$ days and those on the entire network $\mathcal{N}$.}
\label{fig:fig2}
\end{figure}

On the Expo2015 dataset, the ratio between the number of communities identified in the first $n$ days and the communities computed over the whole timespan is greater than $0.90$ and growing for frames larger than $8$ days (Fig.~\ref{fig:fig2}), thus confirming that results are not biased from the length of the selected frame. 
Performance with results obtained for intervals smaller or equal to $8$ days is not optimal but those short intervals are not acceptable because, as shown in Fig.~\ref{fig:fig3}, 1.084 communities (23\%) span periods of $4$ days or longer, which can be not correctly identified with short time scales.

InfoMap identified $4.718$ communities with at least three nodes in $\mathcal{N}$, including $97\%$ of the total number of users.
As expected, most of those communities are very small: ranking them for decreasing number of nodes, the top 643 communities (14\%) include 75\% of all users (see Fig.~\ref{fig:fig1}-(b)).

\begin{figure}[!ht]
\begin{center}
\includegraphics[width=0.7\textwidth]{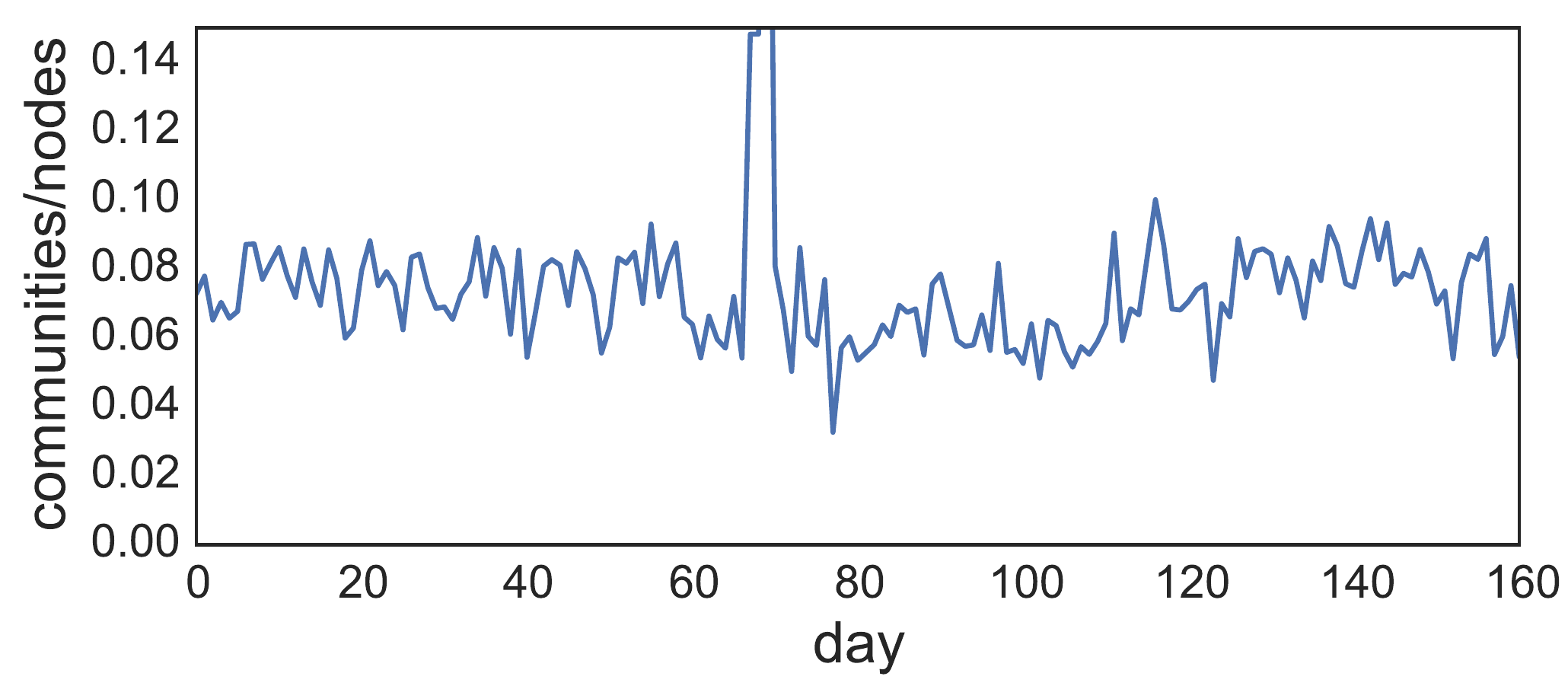}
\end{center}
\caption{Evolution in time of the ratio between the number of detected communities over the number of users.}
\label{fig:fig4}
\end{figure}

The ratio of number of detected communities over the number of users is limited in the interval $[0.04,0.142]$ (Fig.~\ref{fig:fig4}). 
The ratio is indeed nearly constant throughout the whole 5 months period, with the exclusion of a peak period in days [67-69], indeed a technical artefact due to technical issues in collecting tweets for an interval of three days. 

\begin{figure}[!ht]
\begin{center}
\begin{tabular}{cc}
\includegraphics[width=0.5\textwidth]{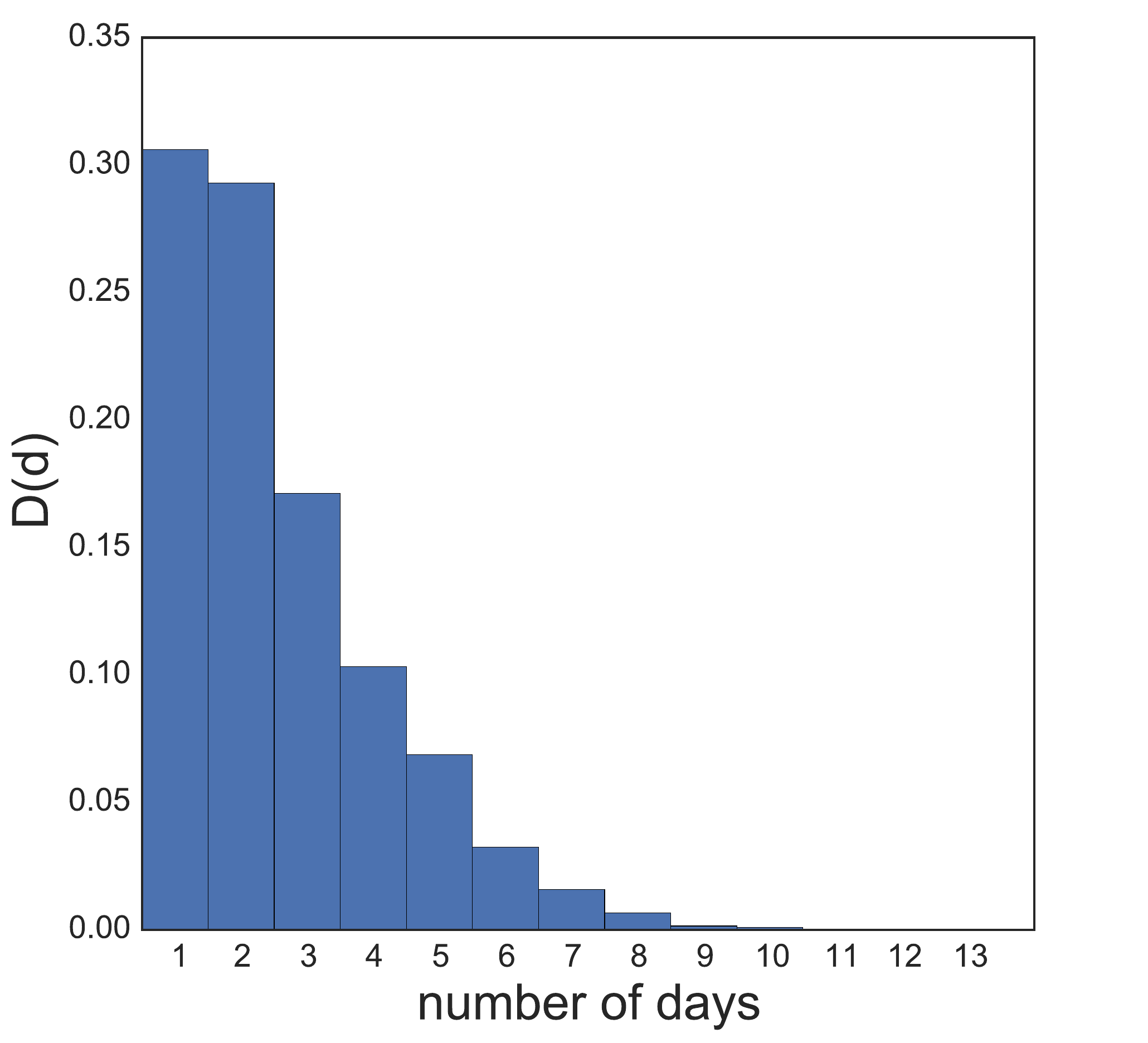} &
\includegraphics[width=0.5\textwidth]{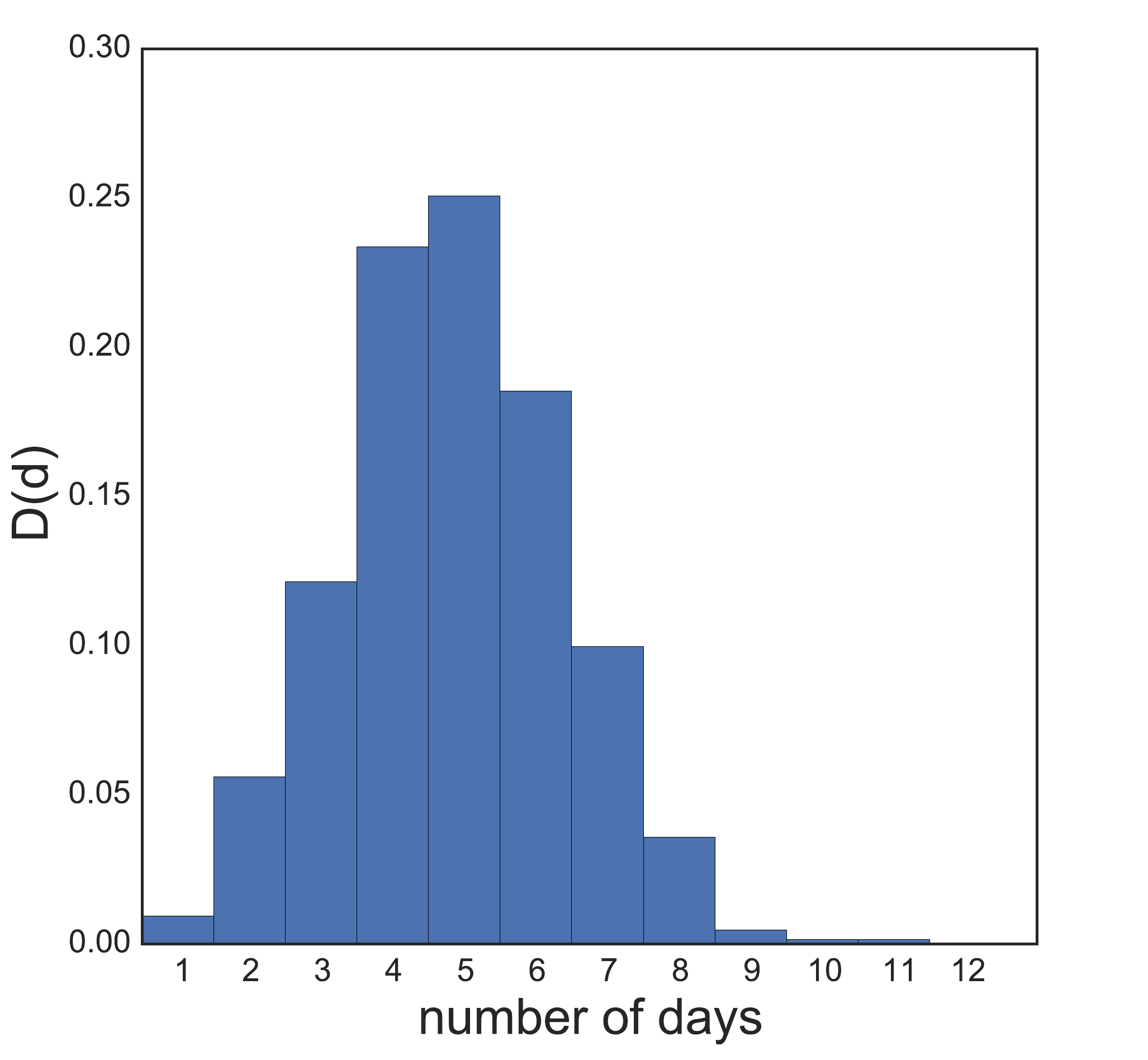} \\
(a) & (b)
\end{tabular}
\end{center}
\caption{Distribution of the community lifespan for all $4.718$ communities (a) and for the subset of the top-643 largest communities (b).}
\label{fig:fig5}
\end{figure}

\begin{figure}[!ht]
\begin{center}
\includegraphics[width=\textwidth]{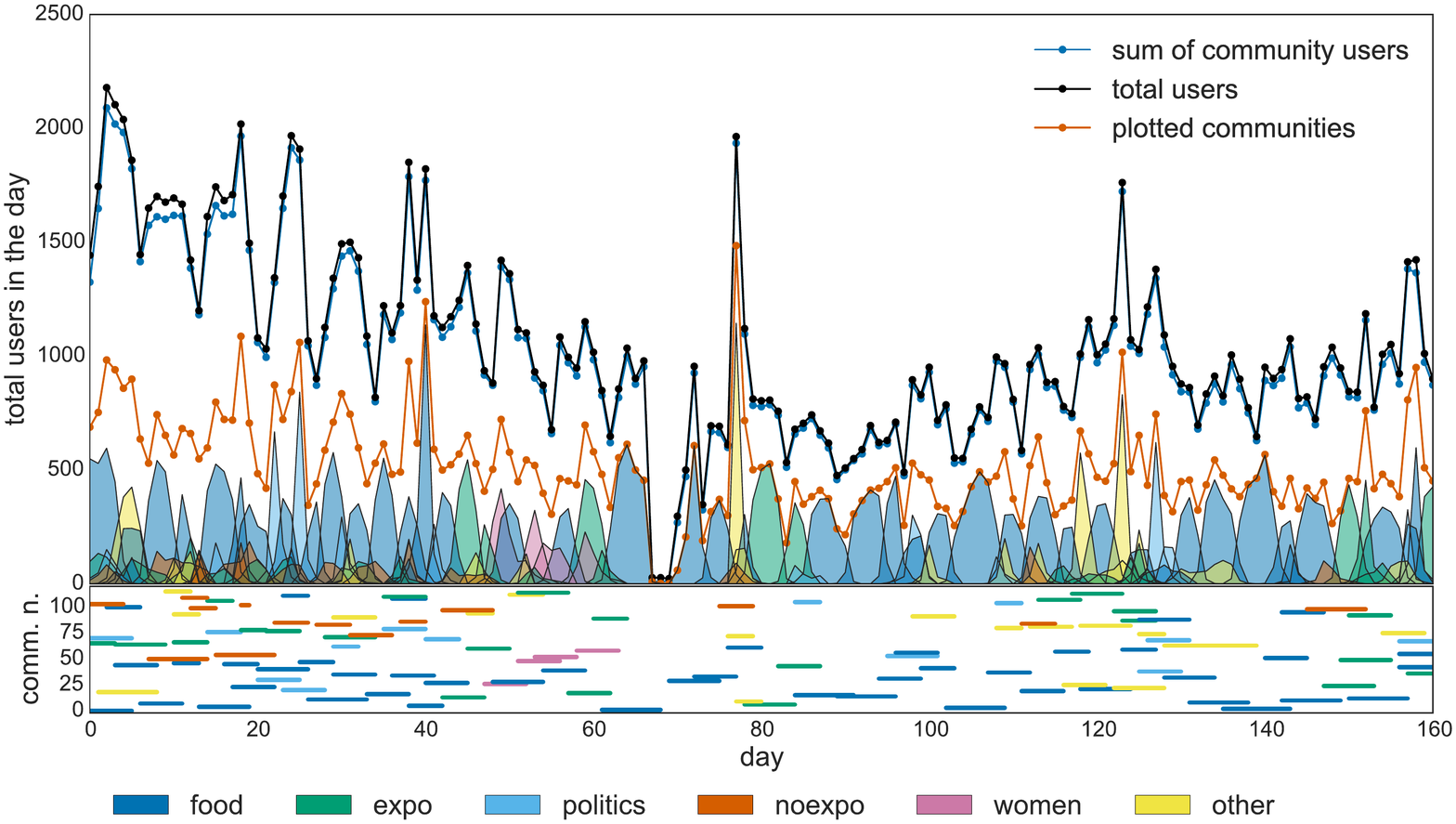}
\end{center}
\caption{Top panel: evolution of the 116 larger communities with the number of corresponding involved users, together with the dynamics of the total number of users per day of $\mathcal{N}$. Lower panel: communities lifetime.}
\label{fig:fig3}
\end{figure}

The distribution of the community lifespan is shown in Fig.~\ref{fig:fig5} (left panel: all $4.718$ communities; right panel: subset of the top-$643$ largest communities). 
Highest $D(d)$ are found for communities covering three or four days at most, or even one-day sets. 
Otherwise, the distribution of top $14\%$ large communities is a Gaussian peaked at 5 days.

A global overview on Expo2015 communities is displayed in Fig.~\ref{fig:fig3}, where the daily number of users is shown for the $116$ larger communities, corresponding to 50\% of the total number of users, with the bottom panel highlighting the community timespans. 
Colors in the plots are chosen according to the three more used hashtags in the community and identifying $6$ main topics (\textit{food}, \textit{expo}, \textit{politics}, \textit{noexpo}, \textit{women} and \textit{other}) from these hashtags.
The artefact due to a gap in data collection is confirmed by the sudden drop to only 30 users around day 68.

Remarkably, a same topic can be shared by two communities overlapping across some days, with one community shrinking while the other is growing.
Even if the topic is the same, the Infomap algorithm tends to differentiate groups rather than merging them: this effect is likely due to the fact that hashtags referring to that topic change with time. 
As discussions in the social networks are expected to reflect specific events, our method picks up naturally growth and disappearing of such communities. 

We have noted that different communities can discuss simultaneously the same topic, but using different hashtags, each one with its own trajectory, thus remaining separated. 
For Expo2015 this clearly happened for the ``\textit{food}'' topic, central to the event and for general comments grouped under ``\textit{expo}'' mostly related on the experience on visiting the exhibition. 
Following these trajectories results transparent in our approach but it is quite hard or even impossible without considering inter-layer links. 
In fact, in the multiplex approach the number of nodes are fixed, therefore the full network results much smaller given that the majority of the people post only one tweet; moreover, setting a connection between communities in different layers is not trivial.

\section{Conclusion}
\label{sec:conclusion}
In this paper we introduced connected multilayer time-dependent networks, where the single snapshot can have different nodes and connection between consecutive layers is described by a matrix. 
As an application, we run a community detection on the connected time-dependent network built from Twitter data related to the Expo2015 Universal Exhibition. 
With this approach, details of the community dynamics can be detected that are hard to identify by the standard approaches. 
The method presented has a general purpose and can be applied to different kind of complex systems, not limited to the analysis of social media data.


\bibliographystyle{unsrt}
\bibliography{cristoforetti15community}
\end{document}